\documentclass[a4paper,oneside,12pt]{article}

\usepackage{epsfig,array,amsmath,amssymb,psfrag,graphicx,tensor}

\newcommand{\beq}{\begin{equation}}
\newcommand{\eeq}{\end{equation}}
\newcommand{\bea}{\begin{eqnarray}}
\newcommand{\eea}{\end{eqnarray}}

\newcommand{\ZZ}{\mathbb{Z}}

\newcommand{\mc}[1]{\mathcal{#1}}

\newcommand{\ii}{\mathrm{i}}
\newcommand{\ee}{\mathrm{e}}

\makeatletter \@addtoreset{equation}{section} \makeatother 
\renewcommand{\theequation}{\arabic{section}.\arabic{equation}}

\begin{document}

\title{Weak cosmic censorship, dyonic Kerr--Newman black holes and Dirac fields}

\author{
G\'abor Zsolt T\'oth
\\[4mm] 
\small \textit{Institute for Particle and Nuclear Physics, Wigner RCP,} \\
\small \textit{MTA Lend\"ulet Holographic QFT Group, Konkoly Thege Mikl\'os \'ut 29-33,} \\
\small \textit{1121 Budapest, Hungary} \\
\small \texttt{toth.gabor.zsolt@wigner.mta.hu} \\
\date{}
}
\maketitle

\begin{abstract}
It was investigated recently, with the aim of testing the weak cosmic censorship conjecture, whether
an extremal Kerr black hole can be converted into a naked singularity by interaction with 
a massless classical Dirac test field, and it was found that this is possible. 
We generalize this result to electrically and magnetically charged rotating extremal black holes
(i.e.\ extremal dyonic Kerr--Newman black holes)
and massive Dirac test fields,
allowing magnetically or electrically uncharged or nonrotating black holes
and the massless Dirac field as special cases.
We show that the possibility of the conversion 
is a direct consequence of the fact that
the Einstein--Hilbert energy-momentum tensor of the classical Dirac field 
does not satisfy the null energy condition, 
and is therefore not in contradiction with the weak cosmic censorship conjecture.
We give a derivation of the absence of superradiance of the Dirac field without making use of the 
complete separability of the Dirac equation in dyonic Kerr--Newman background, 
and we determine the range of superradiant frequencies of the scalar field. 
The range of frequencies of the Dirac field that can be used to convert a black hole into a 
naked singularity partially coincides with the superradiant range of the scalar field.
We apply horizon-penetrating coordinates,
as our arguments involve calculating quantities at the event horizon.
We describe the separation of variables for the Dirac equation in these coordinates, 
although we mostly avoid using it.
\end{abstract}



\maketitle

\thispagestyle{empty}

\newpage

\section{Introduction}
\label{sec.intr}

The well-known weak cosmic censorship conjecture (WCCC), 
stated originally by Penrose \cite{Penrose}, asserts that 
naked singularities (i.e.\ gravitational singularities not hidden behind an event horizon) 
generically cannot be produced in a physical process from regular initial conditions, 
if the matter involved in the process has reasonable properties. 
Although there is significant evidence in favour of the validity of this conjecture, 
finding a general proof
remains one of the major unsolved problems of classical general relativity. 
(For a more detailed and precise description of the WCCC and for reviews on results 
regarding its validity see \cite{WaldGR,Wald0,Joshi,Clarke,Singh,Krolak}.)

As long as a complete proof is not available, 
it is interesting to test the WCCC in various special cases.
A possible such test is a thought experiment in which 
a small particle is thrown at a Kerr--Newman black hole and 
it is checked if an overextremal Kerr--Newman spacetime, 
which contains a naked singularity, 
can arise after the particle has been absorbed by the black hole.
This thought experiment was considered first in \cite{Wald2}, 
where it was shown that an
extremal Kerr--Newman black hole cannot be overcharged or overspun by throwing a 
pointlike test particle with electric charge into it. 
In particular, it was shown that if a particle has a charge 
or angular momentum that would make the black hole 
overextremal if it absorbed the particle, 
then the particle will not fall into the black hole.
A simpler derivation of this result was given in \cite{Needham}. 
In \cite{Hiscock} and \cite{Semiz2} the result of \cite{Wald2} was extended 
to dyonic Kerr--Newman black holes, 
which are rotating black holes with both electric and magnetic charge.
More recently another version of the thought experiment 
in which various test fields (scalar, electromagnetic and Dirac)
are used instead of point particles was also considered \cite{Semiz1, Duztas2, DS, Duztas1, TG}. 
It was found that the weak cosmic censorship is not violated in these cases either, 
with the exception of the case when the test field is a Dirac field \cite{Duztas1}.
Such a result is not surprising, since the WCCC is expected to be valid 
only for matter that has ``reasonable'' properties, 
among which a suitable energy condition is included (see e.g.\ \cite{WaldGR,Wald0}), 
and the Dirac field is well known not to satisfy the weak energy condition \cite{Ch}, 
in contrast with the scalar and electromagnetic fields. 
Studying the case of Dirac test fields is interesting, nevertheless, 
because fermionic matter has an important role in physics.   

In the present paper we extend the result of \cite{Duztas1}, 
which applies to Kerr black holes and massless neutral Dirac fields, 
to charged rotating black holes and charged massive Dirac fields. 
For the sake of generality we allow 
the black hole to have magnetic charge as well, i.e.\ we consider 
dyonic Kerr--Newman black holes,
but we stress that the cases of Kerr--Newman, Reissner--Nordstr\"om and Kerr
black holes and neutral or massless Dirac fields can be obtained from the general case 
by suitable special choice of the parameters.  

The arguments in this paper are technically different from \cite{Duztas1} in a few aspects. 
First, we make little use of the complete separability 
of the Dirac equation in dyonic Kerr--Newman background; we mainly use only 
Fourier expansion in the time and azimuthal angle variables, 
along with simple properties of the Dirac field. 
Second, we apply horizon-penetrating coordinates, 
since these are well suited for calculating fluxes at the event horizon. 
Third, instead of the Newman--Penrose formalism we use 
orthonormal tetrads and four-component Dirac spinor formalism. 
This is done to keep the formalism close to the usual Minkowski spacetime
formulation of Dirac fields (see e.g.\ \cite{PS}).
Fourth, we construct the energy and angular momentum currents 
using Noether's theorem rather than the Einstein--Hilbert energy-momentum tensor, 
because the latter method is not suitable in the presence of
external electromagnetic fields.

The Dirac field has another remarkable feature in which it differs from 
the scalar and electromagnetic fields, namely
it does not exhibit superradiance in black hole spacetimes. 
After discussing the thought experiment 
we present a derivation of this result as well, 
because it requires arguments similar to those used for the thought experiment, 
and because the derivations that can be found
in the literature usually apply the complete separability 
of the Dirac equation (see e.g.\ \cite{Ch,sr,DD,Lee,Unruh:1973,Gueven:1977,Casals:2012es}), 
but we would like to emphasize that this is not necessary. Our derivation is similar to 
the one that is outlined in \cite{WaldGR,Arderucio:2014oua}.
Moreover, the non-superradiant nature of the Dirac field is also related to 
its property that it does not satisfy the weak energy condition (see e.g.\ \cite{Ch,sr,WaldGR}).
We determine the superradiant frequency range of the scalar field as well, 
because it has relevance for the thought experiment. 
The superradiance of the scalar field is discussed 
in several articles (see e.g.\ \cite{WaldGR,sr,RWPU}), 
but usually at zero magnetic charge, and often in a way that 
relies on the complete separability of the field equation.

The paper is organized as follows.
In Section \ref{sec.diraceq} the Dirac field is introduced 
and its conservation laws 
relevant for the thought experiment are discussed. 
This is done in a general setting, 
i.e.\ the discussion is not specialized to black hole spacetimes. 
In Section \ref{sec.knd} the relevant properties of dyonic Kerr--Newman black holes are recalled.  
In Section \ref{sec.ft} the thought experiment is described and the derivation of the main result, 
which indicates
the possibility of the formation of a naked singularity as a result of the interaction
of a black hole and a classical Dirac field, is presented. A discussion of the relevance of backreaction
effects is also included. 
In Section \ref{sec.srad} the absence of superradiance of Dirac fields around 
dyonic Kerr--Newman black holes is derived and
the superradiant frequency range of the scalar field is determined.
Conclusions are given in Section \ref{sec.concl}.
In \ref{sec.A} a part of the formalism of spinor fields in curved spacetime 
is recalled for completeness and to fix notation. 
In \ref{sec.B} the separation of variables for the Dirac equation, 
pertaining to the horizon-penetrating coordinates and to the tetrad used in this paper, is described. 
The asymptotic behaviour of the radial functions at the event horizon is also determined. 

The signature of metric tensors will be $(+,-,-,-)$.

\section{The Dirac field}
\label{sec.diraceq}

The Lagrangian density of the Dirac field $\Psi$ in 
fixed gravitational and electromagnetic fields is 
\beq
\label{eq.lagr}
\mc{L}=
\frac{1}{2} g^{\mu\nu}[\bar{\Psi}\ii\gamma_\mu (\nabla_\nu+\ii e A_\nu) \Psi - (\nabla_\nu-\ii e A_\nu) \bar{\Psi} \ii\gamma_\mu \Psi ] - m\bar{\Psi}\Psi\ ,
\eeq
where $A_\mu$ is the vector potential of the electromagnetic field, 
$m$ is the mass parameter of the Dirac field and $e$ is the electromagnetic coupling constant. 
For the definition of $\nabla_\mu$, $\gamma_\mu$ and $\bar{\Psi}$ see \ref{sec.A}.
The Euler--Lagrange equation corresponding to $\mc{L}$ is the Dirac equation,
$\ii \gamma^\mu (\nabla_\mu +\ii e A_\mu)\Psi=m\Psi$.

\subsection{Conserved currents}
\label{sec.dcons}

The electric current of the Dirac field is 
\beq
j_{em}^\mu=-e\bar{\Psi}\gamma^\mu \Psi=\frac{\partial\mc{L}}{\partial A_\mu}\ .
\eeq 
The closely related current 
$j^\mu=\bar{\Psi}\gamma^\mu \Psi$
is often called particle number density current. 
The vector $j^\mu$ has the important and well known property that it is real, future directed 
and time-like or null for any Dirac spinor $\Psi$, regardless of the equation of motion. 
Furthermore, one can also verify that
$(\bar{\Psi}\gamma_\mu\Psi)(\bar{\Psi}\gamma^\mu\Psi)=4w^*w$,
where $w=\Psi_1^*\Psi_3+\Psi_2^*\Psi_4$,
thus $j^{\mu}$ is null if and only if $w=0$.
These properties of $j^\mu$ imply that the electric charge of a 
classical Dirac field has a definite sign,
which is the same as the sign of $-e$. 

Regarding conserved currents associated with Killing fields, 
a standard way in general relativity to construct such currents 
is to take $\mathbf{T}^{\mu\nu}K_\nu$, 
where $K^\mu$ is the relevant
Killing vector field and $\mathbf{T}^{\mu\nu}$ is the Einstein--Hilbert energy-momentum tensor 
obtained by the variation of the matter action with respect to the metric.
The conservation of $\mathbf{T}^{\mu\nu}K_\nu$ follows from 
$\nabla_\mu \mathbf{T}^{\mu\nu}=0$ and from the Killing equation. 
Although the Lagrangian density (\ref{eq.lagr}) depends explicitly 
(i.e.\ not only through the metric) on the tetrad field, 
the definition 
of the Einstein--Hilbert energy-momentum tensor can be extended to such cases
(see e.g.\ \cite{W}). 
However, as is well known, in the presence of external 
fields (in particular in the presence of an external electromagnetic field) 
generally $\nabla_\mu \mathbf{T}^{\mu\nu}\ne 0$ and $\mathbf{T}^{\mu\nu}K_\nu$ is not conserved, 
thus one has to find some other way to construct a suitable conserved current.
If the matter action is invariant under the diffeomorphisms generated by the Killing field, 
then Noether's theorem is still available for this purpose.  
In the following we discuss the 
Noether currents of the Dirac field associated with Killing fields, 
and compare them with the currents
$\mathbf{T}^{\mu\nu}K_\nu$.

Let us assume that coordinates are chosen so that 
there is one coordinate function, which we denote by $t$, 
for which $K^\mu=(\partial_t)^\mu$. 
In these coordinates $K^\mu$ generates translations of $t$. 
Let us also assume
that the tetrad (and thus also $\gamma^\mu$) is chosen 
so that it is invariant under $t$-translations. 
In addition, the vector potential of the external electromagnetic field 
is also assumed to be invariant under $t$-translations.
In this case the action of the Dirac field is invariant under $t$-translations, 
and the straightforward application of 
Noether's theorem gives the conserved current
\beq
\mc{E}^\mu  =  \frac{\partial\mc{L}}{\partial_\mu\Psi}\partial_t\Psi +\frac{\partial\mc{L}}{\partial_\mu\bar{\Psi}}\partial_t\bar{\Psi}
-\delta\indices{^\mu_t}\mc{L} 
=   \frac{1}{2}(\ii\bar{\Psi}  \gamma^\mu \partial_t \Psi
 -\ii\partial_t\bar{\Psi} \gamma^\mu\Psi)\ .
\label{eq.nd} 
\eeq
On the right hand side the term $\delta\indices{^\mu_t} \mc{L}$ is omitted 
because $\mc{L}=0$ if $\Psi$ satisfies the Dirac equation.
It is worth noting that $\mc{E}^\mu$ is real, and if $\Psi$ has the $t$-dependence 
$\Psi=\ee^{-\ii\omega t}\psi$, then
$\mc{E}^\mu = \omega j^\mu$.  

If the electromagnetic field or $e$ is zero, then $\mathbf{T}\indices{^\mu_t}$ is also conserved, 
thus it is natural to ask what 
the relation between $\mathbf{T}\indices{^\mu_t}$ and $\mc{E}^\mu$ is in this case.
In the following we show that the answer to this question is that the difference between 
these two currents is a current of the 
form $\nabla_\nu f^{\mu\nu}$, where $f^{\mu\nu}$ is antisymmetric, 
therefore  $\mathbf{T}\indices{^\mu_t}$ and $\mc{E}^\mu$ 
can be considered to be equivalent.
In fact we derive a more general result, equation (\ref{eq.zero}),
which holds also in the presence of electromagnetic field. 
(\ref{eq.zero}) will be useful in Section \ref{sec.ft}. 

The Einstein--Hilbert energy-momentum tensor of the Dirac field is
\bea
\mathbf{T}^{\mu\nu} & = &
\frac{1}{4}\Big(\bar{\Psi}\ii\gamma^\mu (\nabla^\nu+\ii e A^\nu) \Psi 
+ \bar{\Psi}\ii\gamma^\nu (\nabla^\mu +\ii e A^\mu)\Psi \nonumber \\
&& \hspace{6mm} -(\nabla^\mu-\ii e A^\mu) \bar{\Psi}\ii \gamma^\nu\Psi 
-(\nabla^\nu-\ii e A^\nu) \bar{\Psi}\ii \gamma^\mu\Psi\Big)\ .
\label{eq.ehd}
\eea
We also introduce the similar tensor 
\beq
\label{eq.tmn}
\hat{T}_{\mu\nu}=
 \frac{1}{2}\Big(\bar{\Psi} \ii \gamma_\mu (\partial_\nu+\ii e A_\nu) \Psi
 -(\partial_\nu-\ii e A_\nu )\bar{\Psi}\ii \gamma_\mu\Psi\Big)\ ,
\eeq
which will appear in Section \ref{sec.ft} as well, and
we define $f^{\mu\nu}$ as
\beq
f^{\mu\nu}  =  -\frac{1}{8}\ii\bar{\Psi} (\gamma^\mu \gamma_t \gamma^\nu- \gamma^\nu\gamma_t\gamma^\mu  )\Psi\ . 
\eeq
By evaluating $\nabla_\nu f^{\mu\nu}$ one finds that if $\Psi$ satisfies the Dirac equation, then   
\bea
\nabla_\nu f^{\mu\nu} & = & \frac{1}{4} [ \ii\bar{\Psi} \gamma^\mu \nabla_t \Psi - \ii\bar{\Psi}\gamma_t \nabla^\mu \Psi
-\ii\nabla_t \bar{\Psi} \gamma^\mu \Psi +\ii\nabla^\mu \bar{\Psi} \gamma_t \Psi ] \nonumber \\ 
&& -
\frac{1}{2} [\ii \bar{\Psi}\gamma^\mu (\nabla_t -\partial_t)\Psi
-
\ii (\nabla_t-\partial_t)\bar{\Psi} \gamma^\mu \Psi]\nonumber \\
&& +\frac{1}{2}eA^\mu\bar{\Psi}\gamma_t\Psi - \frac{1}{2}eA_t\bar{\Psi}\gamma^\mu \Psi\ .
\eea
From this result and from (\ref{eq.ehd}) and (\ref{eq.tmn}), it can be seen immediately that 
\beq
\label{eq.zero}
\hat{T}\indices{^\mu_t}-\mathbf{T}\indices{^\mu_t}= \nabla_\nu f^{\mu\nu}\ .
\eeq
The current on the right hand side is conserved for arbitrary $\Psi$, because $f^{\mu\nu}$ is by definition antisymmetric.

By applying Stokes's theorem it is easy to show, and is well known, that
if a current has the form $\nabla_\nu f^{\mu\nu}$, 
where $f^{\mu\nu}$ is antisymmetric, 
then any corresponding charge associated with some hypersurface 
(which does not need to be space-like) is zero 
if the surface integral arising in the application of Stokes's theorem vanishes. 
Therefore in view of (\ref{eq.zero}) $\hat{T}\indices{^\mu_t}$ 
and $\mathbf{T}\indices{^\mu_t}$ can be considered to be equivalent.

In the absence of electromagnetic field $\hat{T}\indices{^\mu_t} = \mc{E}^\mu$, 
thus in this case (\ref{eq.zero}) shows that  
$\mc{E}^\mu$ and $\mathbf{T}\indices{^\mu_t}$ are equivalent. 

We note that a similar but more special result on the equivalence of $\mc{E}^\mu$ and $\mathbf{T}\indices{^\mu_t}$ 
can be found in \cite{Unruh}.

\section{The dyonic Kerr--Newman black holes}
\label{sec.knd}

A dyonic Kerr--Newman black hole can be characterized by four parameters, 
the mass $M$, the angular momentum per unit mass
$a$, the electric charge $Q_e$ and the magnetic charge $Q_m$. 
The angular momentum of the black hole is $J=aM$, and
$Q_m=0$ corresponds to a usual Kerr--Newman black hole. 
The metric of the dyonic Kerr--Newman black hole spacetime with parameters $(M,a,Q_e,Q_m)$
is the same as the Kerr--Newman metric with parameters $(M,a, q)$, $q^2 = Q_e^2+Q_m^2$, where $q$ denotes the electric charge parameter 
of the Kerr--Newman metric.
The parameters have to satisfy the inequality 
\beq
\label{eq.eta}
\eta=M^2-Q_e^2-Q_m^2-a^2\ge 0\ ,
\eeq
otherwise the spacetime contains a naked singularity. The black hole is called extremal if 
$\eta=0$. 
Under certain conditions, the dyonic Kerr--Newman black holes are the only static and asymptotically flat black hole solutions
of the Einstein--Maxwell equations \cite{Mazur,Bunting}.

The vector potential of the electromagnetic field of a dyonic Kerr--Newman black hole is 
\beq
A=Q_e A_e+ Q_m A_m\ , 
\eeq 
where 
\bea
\label{eq.ae}
A_e & = & -\frac{r}{\Sigma}dt  +  \frac{a r \sin^2\theta}{\Sigma}d\phi\\
\label{eq.am}
A_m & = & \frac{ a\cos\theta}{\Sigma}dt  +
\left[\tilde{C}-  \frac{r^2+a^2}{\Sigma}\cos\theta  \right]   d\phi\ ,
\eea
\beq
\Sigma  =  r^2 + a^2 \cos^2\theta\ .
\eeq
These formulas are written in Boyer-Lindquist coordinates $(t,r,\theta,\phi)$.
The electromagnetic field derived from $A_m$ is dual to the electromagnetic field derived from $A_e$.
The electromagnetic field does not depend on the constant $\tilde{C}$,
which can be used, by setting $\tilde{C}=1$ or $\tilde{C}=-1$, to eliminate 
the Dirac string singularity of $A_m$ along the positive or negative $z$ axis ($\theta=0$ and $\theta=\pi$), respectively. We set $\tilde{C}$ to zero for a reason that is explained below.

\subsection{Horizon-penetrating coordinates}
\label{sec.hp}

In the following sections various quantities will be considered at the future event horizon. 
Since the Boyer--Lindquist coordinates do not cover the future event horizon,   
Eddington--Finkelstein-type ingoing horizon-penetrating coordinates, 
denoted by $(\tau,r,\theta,\varphi)$, will be used. 
These coordinates can be introduced by the transformation
\beq
\tau  =  t-r+\int dr\, \frac{r^2+a^2}{\Delta}\ ,\qquad 
\varphi = \phi+ \int dr\, \frac{a}{\Delta} \ ,
\eeq
where
$\Delta  =  r^2 + a^2 + Q_e^2 +Q_m^2 -2Mr$.
The future event horizon is located in these coordinates at the constant value 
$r_+=M+\sqrt{M^2-(a^2+Q_e^2+Q_m^2)}$ 
of $r$, and the metric is non-singular in these points. The inner horizon is located at
$r_-=M-\sqrt{M^2-(a^2+Q_e^2+Q_m^2)}$.  
In the extremal case  
$r_+=r_-=M$.
The $(\tau+r, \theta,\varphi) = constant$ lines are ingoing null geodesics, and 
there exists an $r_0< r_+$ such that 
the $\tau=constant$ hypersurfaces
are space-like in the domain $r_0< r$. 

The $r$ component $(A_e)_r$ of $A_e$ with respect to the coordinates $(\tau,r,\theta,\varphi)$
is singular at the event horizon, but this singularity can be eliminated by 
the 
gauge transformation $A_e \to A_e - \frac{r}{\Delta} dr$. 
After this gauge transformation 
\beq
\label{eq.ae2}
A_e=-\frac{r}{\Sigma}d\tau  +  \frac{a r \sin^2\theta}{\Sigma}d\varphi
-\frac{r}{\Sigma} dr\ .
\eeq

The $r$ component of $A_m$ with respect to the coordinates $(\tau,r,\theta,\varphi)$ is also singular 
if $\tilde{C}\ne 0$, therefore we set $\tilde{C} = 0$. 
Nevertheless, in order to treat the Dirac string singularity of $A_m$, 
we introduce an explicit gauge parameter into it by adding $Cd\varphi$, where $C$ is a real constant. Thus  
\beq
\label{eq.am2}
A_m=\frac{ a\cos\theta}{\Sigma}d\tau  +
\left[C-   \frac{r^2+a^2}{\Sigma}\cos\theta \right]   d\varphi
+\frac{a\cos\theta}{\Sigma}dr\ .
\eeq
Generally $A_m$ has a string singularity along 
the $z$ axis (which corresponds to $\theta=0$ and $\theta=\pi$) 
because $d\varphi$ is singular here, and its coefficient $(A_m)_\varphi$
does not cancel this singularity. 
However, in the special cases $C=1$ and $C=-1$ the singularity is cancelled
along the positive $z$ axis ($\theta=0$) or along the negative $z$ axis ($\theta=\pi$), respectively.
The string singularity can therefore be avoided by using 
two domains that cover the whole spacetime region of interest 
in such a way that one of the domains contains 
the entire positive $z$ axis but is well separated from 
the negative $z$ axis and the other one contains 
the entire negative $z$ axis but is separated from the 
positive $z$ axis. In the first domain the $C=1$ gauge is used then, 
and in the second domain the $C=-1$ gauge. 
Suitable domains are given by the relations 
$r_0<r$, $0\leq \theta < \pi/2+\epsilon$ and $r_0<r$, $\pi/2-\epsilon < \theta \leq \pi$, 
where $\epsilon$ is some small number. 
These domains will be denoted by $\mc{D}_+$ and $\mc{D}_-$. 
It should be kept in mind that 
the transition between the two domains involves a gauge transformation.
This approach to treating the string singularity of $A_m$ 
was proposed in \cite{WY} and was taken also in \cite{Semiz2,Semiz1,TG}.    

In the following sections and in \ref{sec.B}, except in Section \ref{sec.tetrad},
we use only the coordinates $(\tau,r,\theta,\varphi)$, 
and we also use the notation $\zeta$ for the one-form 
$dr$ (the exterior derivative of the coordinate function $r$), i.e.\
\beq
\zeta^\mu = (dr)^\mu\ .
\eeq
$A_e$, $A_m$ and $A$ will denote (\ref{eq.ae2}), (\ref{eq.am2}) and $A=Q_e A_e+ Q_m A_m$, respectively.

\subsection{Various important properties}
\label{sec.vip}

In this section further important properties of the Dyonic Kerr--Newman black holes,
which will be used in the subsequent sections, are collected. 

$\partial_\tau$ and $\partial_\varphi$ are Killing fields; 
$\partial_\tau$ is the generator of time translations and $\partial_\varphi$ is the generator of rotations around the 
axis of the black hole.
$A_e$ and $A_m$ are also invariant under 
these symmetries.
The Killing field
\beq
\label{eq.chi}
\chi=\partial_\tau +\Omega_H \partial_\varphi\ ,\qquad \Omega_H=\frac{a}{r_+^2+a^2}
\eeq
is null at the event horizon.
In the subsequent sections it will also be important that at the event horizon
\beq
\label{eq.a}
(A_e)_\mu\chi^\mu=\frac{-r_+}{r_+^2+a^2}\ ,\qquad
(A_m)_\mu\chi^\mu=C\Omega_H\ ,
\eeq
and  $\zeta^\mu$ is parallel to $\chi^\mu$. The relation between $\zeta^\mu$ 
and $\chi^\mu$ at the event horizon is
\beq
\label{eq.rho}
\zeta^\mu=-\frac{r_+^2+a^2}{r_+^2+a^2\cos^2\theta} \chi^\mu\ ,
\eeq
thus $\zeta^\mu$ is past directed (in \cite{TG} $\zeta^\mu$ was denoted by $\omega^\mu$ and it was future directed because of the opposite signature of the metric there).
(\ref{eq.rho}) shows that  $\zeta^\mu$ is null at the event horizon, 
but it should be stressed that this property of $\zeta^\mu$ follows directly from the facts that 
the event horizon is a null surface and is a level surface of the function $r$. 

It is useful to introduce the quantity $\Phi_H$ as
\beq
\label{eq.h}
\Phi_H=\frac{r_+ Q_e}{r_+^2+a^2}\ .
\eeq
In the case of Kerr--Newman black holes, $\Phi_H$ is known as the electrostatic potential of the horizon.

\subsection{Tetrad}
\label{sec.tetrad}

In order to define a suitable tetrad for the Kerr--Newman metric one can start with the 
Kinnersley-type tetrad (see also \cite{Semiz3}) 
\bea
V^{\bar{0}}_\mu & = & \frac{1}{\sqrt{2}}\left(
\left(1 + \frac{\Delta}{2\Sigma}\right) dt+   
\left(\frac{1}{2} - \frac{\Sigma}{\Delta}\right) dr 
-\left(1 + \frac{\Delta}{2\Sigma}\right)a\sin^2\theta\, d\phi
\right)\\
V^{\bar{1}}_\mu & = & 
-\frac{a^2\cos\theta\sin\theta}{\Sigma}dt
+r\, d\theta
+\frac{a(a^2+r^2)\cos\theta\sin\theta}{\Sigma}d\phi \\
V^{\bar{2}}_\mu & = & 
\frac{ar\sin\theta}{\Sigma}dt
+a\cos\theta\, d\theta
-\frac{r(a^2+r^2)\sin\theta}{\Sigma}d\phi \\
V^{\bar{3}}_\mu & = &  \frac{1}{\sqrt{2}}\left(
\left(-1+\frac{\Delta}{2\Sigma}\right) dt
+\left(\frac{1}{2}+\frac{\Sigma}{\Delta}\right)dr
+\left(1-\frac{\Delta}{2\Sigma}\right)a\sin^2\theta\, d\phi
\right),
\eea
given in Boyer--Lindquist coordinates. This can be transformed into the
ingoing horizon-penetrating coordinates, but one finds that it is singular at the event horizon.
This singularity can nevertheless be removed by a suitable local Lorentz transformation, 
similarly as for example in \cite{Teukolsky}. 
Thus in the present paper we use the Lorentz transformed non-singular tetrad $\tilde{V}_\mu^{\bar{\mu}}$ 
related to $V_\mu^{\bar{\mu}}$ as  
\bea
\tilde{V}^{\bar{0}}_\mu & = & 
\frac{r^2}{\Delta} (V^{\bar{0}}_\mu + V^{\bar{3}}_\mu)
+ \frac{\Delta}{r^2} (V^{\bar{0}}_\mu - V^{\bar{3}}_\mu)\\
\tilde{V}^{\bar{3}}_\mu & = & 
\frac{r^2}{\Delta} (V^{\bar{0}}_\mu + V^{\bar{3}}_\mu)
- \frac{\Delta}{r^2} (V^{\bar{0}}_\mu - V^{\bar{3}}_\mu)
\eea
\beq
\tilde{V}^{\bar{1}}_\mu = V^{\bar{1}}_\mu\, ,\qquad
\tilde{V}^{\bar{2}}_\mu = V^{\bar{2}}_\mu\ .
\eeq
$\tilde{V}_\mu^{\bar{\mu}}$ and $V_\mu^{\bar{\mu}}$ are invariant under time translations and under rotations around the axis of the 
black hole, and $\tilde{V}_\mu^{\bar{\mu}}$ tends to $V_\mu^{\bar{\mu}}$ if $r\to\infty$. 
It should also be mentioned that $V_\mu^{\bar{\mu}}$ and $\tilde{V}_\mu^{\bar{\mu}}$ are not null tetrads, rather
$V_\mu^{\bar{\mu}}V^{\mu\bar{\nu}}= \tilde{V}_\mu^{\bar{\mu}}\tilde{V}^{\mu\bar{\nu}}=g^{\bar{\mu}\bar{\nu}}$, where
$g^{\bar{\mu}\bar{\nu}}=\mathrm{diag}(1,-1,-1,-1)$. We note finally that
another useful choice for $V_\mu^{\bar{\mu}}$ would be the `canonical' tetrad of Carter \cite{Semiz3,Carter}.

\section{The thought experiment}
\label{sec.ft}

The thought experiment for testing the WCCC is assumed to proceed in the following way. Initially 
one has an extremal dyonic Kerr--Newman black hole, 
then a small amount of matter represented by a wave packet is thrown at it from great distance. A certain part of the 
matter is absorbed by the black hole, the remaining part is scattered back to infinity, 
and finally the system settles down in another 
dyonic Kerr--Newman state with slightly different parameters. 

Under an infinitesimally small change $(dM,dJ,dQ_e,dQ_m)$ of the parameters\\ 
$(M,J,Q_e,Q_m)$ of a dyonic Kerr--Newman configuration
the change of $\eta$ (which was introduced in (\ref{eq.eta})) is 
\beq
\label{eq.deta}
d\eta=2\frac{M^2+a^2}{M}\left( dM-\frac{a}{M^2+a^2}dJ-\frac{Q_e M}{M^2+a^2}dQ_e-\frac{Q_m M}{M^2+a^2}dQ_m  \right).
\eeq
If one calculates the change $(dM,dJ,dQ_e,dQ_m)$ of the parameters 
in the process described above, 
one should find $d\eta\ge 0$, if the final state is a dyonic Kerr--Newman black hole and
cosmic censorship is not violated, 
whereas a result $d\eta < 0$ indicates the formation of a naked singularity, 
and thus a violation of the WCCC. Of course, in the case $d\eta < 0$ the last conclusion that 
the WCCC is violated can be drawn only if the matter used in the thought experiment 
does have the properties required in the WCCC.

In the calculation of $(dM,dJ,dQ_e,dQ_m)$ the test matter approximation is used, 
i.e.\ the metric and the electromagnetic field 
are considered fixed and backreaction effects are neglected. 
The reason for taking the initial black hole state to be extremal 
is that the quantities $(dM,dJ,dQ_e,dQ_m)$ are very small, 
in accordance with the test matter approximation.  

There are several articles, e.g.\ \cite{Hubeny}-\cite{Duztas:2015oja}, 
in which other versions or aspects of the 
thought experiment are studied. For instance, backreaction effects and 
subextremal initial black holes are considered in several papers. 
Furthermore, besides the thought experiment 
it is interesting to study the possibilities of 
observing naked singularities that may form if the WCCC is violated; 
see e.g.\ \cite{PQR}-\cite{Ortiz:2015rma}.

\vspace{3mm}
We turn now to the calculation of $d\eta$. 
In the following the black hole is not restricted to be extremal unless explicitly stated.
Applying (\ref{eq.nd}) to the Killing fields $(\partial_\tau)^\mu$ and $(\partial_\varphi)^\mu$
one obtains that the energy and angular momentum currents are given by the equations
\bea
\label{eq.energy}
\mc{E}^\mu & = & \hat{T}\indices{^\mu_\tau} + e A_\tau j^\mu\\
\label{eq.angmom}
\mc{J}^\mu & = & \hat{T}\indices{^\mu_\varphi} + e A_\varphi j^\mu\ ,
\eea
where $\hat{T}_{\mu \nu}$ is given by (\ref{eq.tmn}).

$\hat{T}_{\mu\nu}$ and $j^\mu$ are gauge invariant and $A_\tau$ does not depend on 
the gauge parameter $C$,
therefore $\mc{E}^\mu$ is also independent of $C$. 
$A_\varphi$ does depend on $C$, however, thus $\mc{J}^\mu$ also depends on it.
For this reason we take (as in \cite{TG}) the modified definition   
\beq
\label{eq.jmod}
\mc{J}^\mu= \hat{T}\indices{^\mu_\varphi} + e(A_\varphi - Q_m C)j^\mu 
\eeq
for $\mc{J}^\mu$, which eliminates its dependence on $C$. 
The conservation of $\mc{J}^\mu$ is not affected by this modification, 
because $j^\mu$ is conserved.  
The independence of $\mc{E}^\mu$ and $\mc{J}^\mu$ of $C$ is important because 
the value of $C$ is different in the domains $\mc{D}_+$ and $\mc{D}_-$. 

The electric charge flux through the event horizon into the black hole is
\beq
\label{eq.q}
\frac{dQ}{d\tau} =  \int_H \sqrt{-g}\ ej^r\, d\theta d \varphi\ ,
\eeq 
where $H$ denotes the two-dimensional surface of the black hole 
(which is the relevant time slice of the event horizon),
and the energy and angular momentum fluxes are 
\bea
\label{eq.e}
\frac{dE}{d\tau} & = & -\int_H \sqrt{-g}\ \left[{\hat{T}^r }_{\ \, \tau}  + e A_\tau j^r \right]   \, d\theta d \varphi\\
\label{eq.l}
\frac{dL}{d\tau} & = & \phantom{-} \int_H \sqrt{-g}\ \left[{\hat{T}^r}_{\ \, \varphi}   + e  
(A_\varphi-Q_m C)  j^r \right] \, d\theta d \varphi\ ,
\eea
where the quantities in the brackets are $\mc{E}^r$ and $\mc{J}^r$, respectively.
The total energy, angular momentum and electric charge that falls through the event horizon is 
$\int_{-\infty}^\infty  \frac{dE}{d\tau} d\tau$,   
$\int_{-\infty}^\infty   \frac{dL}{d\tau} d\tau  $
and $\int_{-\infty}^\infty  \frac{dQ}{d\tau} d\tau$, respectively. 
The metric and the electromagnetic field are taken to be fixed, 
therefore these quantities can be identified with $dM$, $dJ$ and $dQ_e$, 
i.e.\ with the change of the mass, angular momentum and 
electric charge of the black hole. 
$dQ_m=0$, since the Dirac field does not have magnetic charge.

From the equations (\ref{eq.q}), (\ref{eq.e}), (\ref{eq.l}) above and from (\ref{eq.chi}), (\ref{eq.a}) and (\ref{eq.h})
it follows immediately that 
\beq
\label{eq.tt}
-\int_H \sqrt{-g}\ \hat{T}_{\mu\nu}\zeta^\mu \chi^\nu   \, d\theta d \varphi\
=\
\frac{dE}{d\tau}-\Omega_H  \frac{dL}{d\tau}  -  \Phi_H \frac{dQ}{d\tau}\ .
\eeq
Taking into account the relations $dM=\int_{-\infty}^\infty  \frac{dE}{d\tau} d\tau$,   
$dJ=\int_{-\infty}^\infty   \frac{dL}{d\tau} d\tau  $
and $dQ_e=\int_{-\infty}^\infty  \frac{dQ}{d\tau} d\tau$,
\beq
\label{eq.oo}
-\int_{-\infty}^\infty d\tau \int_H \sqrt{-g}\ \hat{T}_{\mu\nu}\zeta^\mu \chi^\nu   \, d\theta d \varphi
\ 
=\
dM-\Omega_H dJ -  \Phi_H dQ_e
\eeq
is obtained from (\ref{eq.tt}). It is easy to see that in the extremal case the
right hand side in (\ref{eq.oo}) is $\frac{M}{2(M^2+a^2)}d\eta$, 
thus the sign of $d\eta$ depends, in the extremal case, on the sign of
$\int_{-\infty}^\infty d\tau \int_H \sqrt{-g}\ \hat{T}_{\mu\nu}\zeta^\mu \chi^\nu   \, d\theta d \varphi\, $.

In order to examine 
$\int_{-\infty}^\infty d\tau \int_H \sqrt{-g}\ \hat{T}_{\mu\nu}\zeta^\mu \chi^\nu   \, d\theta d \varphi\, $
it is useful to consider the Fourier expansion 
\beq
\label{eq.fourier}
\Psi= \sum_n \int d\omega\ \ee^{-\ii\omega \tau}\ee^{\ii (n- C eQ_m)\varphi}\psi_{\omega,n}(r,\theta)
\eeq
of $\Psi$,
where $\ee^{-\ii\omega \tau}\ee^{\ii (n- C eQ_m)\varphi}\psi_{\omega,n}(r,\theta)$ 
are solutions of the Dirac equation.
The term  $-CeQ_m$ in the factor $\ee^{\ii (n- C eQ_m)\varphi}$ 
is included because of the gauge transformation
done at equation (\ref{eq.am2})  (see also \cite{Semiz3}). 
For $\ee^{-\ii\omega \tau}\ee^{\ii (n- C eQ_m)\varphi}\psi_{\omega,n}(r,\theta)$ 
to be single valued for both $C=1$ and $C=-1$,
both $n- eQ_m$ and $n+ eQ_m$ have to be integer, 
implying that $n$ and $eQ_m$ are either integer or half-integer.
The summation in (\ref{eq.fourier}) should therefore be done 
over $\ZZ$ if $eQ_m$ is integer and over $\frac{1}{2}+\ZZ$ if 
$eQ_m$ is half-integer.
Far from the black hole only modes with $|\omega|>m$ describe propagating waves.
Using (\ref{eq.fourier}) one finds that
\bea
&&\hspace{-1.2cm}\int_{-\infty}^\infty d\tau \int_H \sqrt{-g}\ \hat{T}_{\mu\nu}\zeta^\mu \chi^\nu   \, d\theta d \varphi
\ = \nonumber\\ 
&& \hspace{-0.7cm} = (2\pi)^2 \sum_n \int d\omega\, \int_H d\theta\, \sqrt{-g}\  \zeta^\mu \bar{\psi}_{\omega,n} \gamma_\mu \psi_{\omega,n} \left(\omega - n\Omega_H + \frac{e Q_e r_+}{r_+^2 +a^2} \right).
\label{eq.integr}
\eea 
In the derivation of (\ref{eq.integr}) 
the only derivatives of $\Psi$ that appear are $\partial_\tau\Psi$ and $\partial_\varphi\Psi$, 
which are easy to evaluate,
and formulas (\ref{eq.chi}) and (\ref{eq.a}) can also be applied. 

As was mentioned in Section \ref{sec.vip}, 
$\zeta^\mu$ is a past directed null vector at the event horizon, 
and in Section \ref{sec.dcons} it was also noted that 
$\bar{\psi}_{\omega,n} \gamma_\mu \psi_{\omega,n}$ is a real future directed 
null or time-like vector, therefore $\zeta^\mu \bar{\psi}_{\omega,n} \gamma_\mu \psi_{\omega,n}\le 0$. 
The integrand on the right hand side of  
(\ref{eq.integr}) is thus positive if 
\beq
\label{eq.to}
\tilde{\omega}=\omega -n\Omega_H+ \frac{e Q_e r_+}{r_+^2 +a^2} < 0
\eeq
and  $\zeta^\mu \bar{\psi}_{\omega,n} \gamma_\mu \psi_{\omega,n} \ne 0$ at the event horizon. 
(Here the notation $\tilde{\omega}$ has been introduced.) 
If $\zeta^\mu \bar{\psi}_{\omega,n} \gamma_\mu \psi_{\omega,n}\tilde{\omega}$ 
is large at the event horizon 
mainly for those values of $\omega$ and $n$ for which $\tilde{\omega}<0$, 
then it is possible for the whole integral
(\ref{eq.integr}) to be positive. In this case  
$dM-\Omega_H dJ -  \Phi_H dQ_e < 0$,
in particular in the extremal case $d\eta < 0$, 
indicating a possible violation of the WCCC. 

Clearly $\tilde{\omega}$ is negative if $\omega$ has a sufficiently 
large negative value, but, more interestingly,  
$\tilde{\omega}<0$ is possible even for $\omega>0$, 
if 
\beq
n\Omega_H - \frac{e Q_e r_+}{r_+^2 +a^2} > 0\ .
\eeq
It is also interesting to note 
that the frequency range where $\tilde{\omega}<0$ 
partially coincides with the range where the scalar field exhibits superradiance (see Section \ref{sec.srad2}).

In the special case when the charges $Q_e$ and $Q_m$ of the black hole are zero, 
one can use instead of (\ref{eq.energy}) and (\ref{eq.angmom}) 
the energy and angular momentum currents
obtained from the Einstein--Hilbert energy-momentum tensor, 
as is usually done in the literature (see, for example, \cite{Duztas1,Ch}). 
In view of the arguments in the last part of Section \ref{sec.dcons}, 
this would give the same result 
(namely $-1$ times the right hand side of (\ref{eq.integr}))
for $dM-\Omega_H dJ - \Phi_H dQ_e$. 

A tensor similar to $\hat{T}^{\mu\nu}$ 
appears also in that version of the thought experiment 
in which the test field is a scalar field (see Section 4.1 of \cite{TG} and Section \ref{sec.srad2}).
In that case $\hat{T}^{\mu\nu}$ satisfies the null energy condition
$\hat{T}^{\mu\nu} \chi_\mu \chi_\nu \ge 0$ 
at the event horizon, and 
this implies that the WCCC is not violated. 
If this null energy condition held
in the case of the Dirac test field, 
then it could be used in the same way as in the case of the scalar test field to show
that $dM-\Omega_H dJ -  \Phi_H dQ_e \ge 0$ and the WCCC is not violated. 

In the case of the scalar field $\hat{T}^{\mu\nu}$ 
is the Einstein--Hilbert energy-momentum tensor, 
and it was shown in the last part of Section \ref{sec.dcons}
that $\hat{T}^{\mu\nu}$ is related to the Einstein--Hilbert energy-momentum tensor
also in the case of the Dirac field.
Moreover, 
$\hat{T}_{\mu\nu}\zeta^\mu \chi^\nu = \hat{T}\indices{^r_\tau}+ \Omega_H \hat{T}\indices{^r_\varphi}$,
therefore (\ref{eq.zero}) and Stokes's theorem implies that the left hand side of 
(\ref{eq.oo}) can be written also as
$-\int_{-\infty}^\infty d\tau \int_H \sqrt{-g}\ \mathbf{T}_{\mu\nu}\zeta^\mu \chi^\nu   \, d\theta d \varphi$, where $\mathbf{T}_{\mu\nu}$ is the 
Einstein--Hilbert energy-momentum tensor of the Dirac field given by (\ref{eq.ehd}).
Thus the result (\ref{eq.oo}) for $dM-\Omega_H dJ -  \Phi_H dQ_e$ is completely
analogous to the result obtained in the case of the scalar field in \cite{TG},  
and one can say that the
conversion of a black hole into a naked singularity by a Dirac field is possible
because the Einstein--Hilbert energy-momentum tensor of the Dirac field does not satisfy the 
null energy condition $\mathbf{T}^{\mu\nu} \chi_\mu \chi_\nu \ge 0$. 

The case of combined scalar and electromagnetic test matter was also
considered in Section 4.2 of \cite{TG}, and also in that case it was found that
$dM-\Omega_H dJ -  \Phi_H dQ_e =-\int_{-\infty}^\infty d\tau \int_H \sqrt{-g}\ T_{\mu\nu}\zeta^\mu \chi^\nu   \, d\theta d \varphi$, where $T_{\mu\nu}$ 
is the relevant Einstein--Hilbert energy-momentum tensor. This $T_{\mu\nu}$
satisfies the null energy condition, implying $d\eta \ge 0$. 
If the scalar field vanishes, then this case reduces to the case of purely electromagnetic test field.

The integrand on the right hand side of (\ref{eq.integr}) 
can be expressed in a more explicit form.
At the event horizon 
\beq
\zeta^\mu \gamma^{\bar{0}}\gamma_\mu =\gamma^{\bar{0}}\gamma^r = \frac{-r_+^2}{\sqrt{2}(r_+^2+a^2\cos^2\theta)}\left(
\begin{array}{llll}
1 & 0 & 0 & 0\\
0 & 0 & 0 & 0\\
0 & 0 & 0 & 0\\
0 & 0 & 0 & 1
\end{array} \right)\ ,
\eeq
thus
\beq
\zeta_\mu \bar{\Psi}\gamma^\mu \Psi = \bar{\Psi}\gamma^r \Psi = \frac{-r_+^2}{\sqrt{2}(r_+^2+a^2\cos^2\theta)}\bigl(|\Psi_1|^2+|\Psi_4|^2\bigr)\ ,
\eeq
where $\Psi_1$ and $\Psi_4$ denote the first and fourth components of $\Psi$. Furthermore, 
\beq
\sqrt{-g}\, =\, (r^2+a^2\cos^2\theta)\sin\theta\ ,
\eeq
therefore at the event horizon
\beq
\label{eq.sqrtg}
\sqrt{-g}\, \zeta_\mu \bar{\Psi}\gamma^\mu \Psi = \frac{-r_+^2}{\sqrt{2}}\sin\theta\, \bigl(|\Psi_1|^2+|\Psi_4|^2\bigr)\ .
\eeq
These formulas hold for any spinor $\Psi$, regardless of the Dirac equation, 
thus they hold also for $\psi_{\omega,n}$. (\ref{eq.integr}) can be rewritten therefore as 
\bea
&&\int_{-\infty}^\infty d\tau \int_H \sqrt{-g}\ \hat{T}_{\mu\nu}\zeta^\mu \chi^\nu   \, d\theta d \varphi
\ = \nonumber\\ 
&& \hspace{0.3cm} = (2\pi)^2 \sum_n \int d\omega\, \int_H d\theta\,  
\frac{-r_+^2}{\sqrt{2}}\sin\theta\, \bigl(|(\psi_{\omega,n})_1|^2+|(\psi_{\omega,n})_4|^2\bigr)\,
\tilde{\omega}\ .
\label{eq.integr2}
\eea

Finally, for $d\eta < 0$ it is necessary that $(\psi_{\omega,n})_1$ or $(\psi_{\omega,n})_4$ 
be nonzero at the event horizon 
at least for some values of $\omega$ and $n$ for which $\tilde{\omega}< 0$, 
therefore in principle 
it should be investigated if there is anything 
that could force $(\psi_{\omega,n})_1$ or $(\psi_{\omega,n})_4$ 
to be zero at the event horizon.
If, invoking the separability of the Dirac equation (see \ref{sec.B}), 
it is assumed that $\psi_{\omega,n}$ is a linear combination
of terms satisfying the ansatz (\ref{eq.f}), (\ref{eq.ans1}), (\ref{eq.ans2}), 
then $(\psi_{\omega,n})_1$ or $(\psi_{\omega,n})_4$ is nonzero at the event horizon 
if $R_+(r_+)\ne 0$ in these terms. 
((\ref{eq.f}), (\ref{eq.ans1}) and (\ref{eq.ans2}) show that
$R_-$ does not enter $(\psi_{\omega,n})_1$ and $(\psi_{\omega,n})_4$).)
As explained in more detail in \ref{sec.B.1}, 
$R_+(r_+)$ is not zero,
therefore generally $(\psi_{\omega,n})_1$ and $(\psi_{\omega,n})_4$ 
do not have to be zero at the event horizon.

\subsection{On possible backreaction effects}

Regarding the question whether backreaction effects can be expected to prevent the formation
of a naked singularity, it should be noted first that a result in rigorous test field approximation,
which can be considered as a lowest order approximation, indicating the formation of a naked 
singularity is more conclusive than a result 
which indicates that a naked singularity is not formed (as in the cases of scalar and electromagnetic fields), 
because the latter type of result does not 
exclude the possibility of the formation of a naked singularity outside the domain of validity of the 
test field approximation, whereas the first type of result implies that the formation of a naked singularity 
may be avoided only if the perturbation is sufficiently large so that higher order effects can dominate.
This also shows that considering higher order effects is more important when naked singularity formation is excluded
at lowest order.

In the literature it has been emphasized that backreaction effects have to be taken into account properly, 
and that this usually restores the cosmic censor in scenarios in which it seems to be violated 
\cite{Hod:2008zza, BCK1,BCK2,CB,Colleoni:2015ena}. 
In these cases, in contrast with the case of the Dirac field,
cosmic censorship is respected at lowest order, i.e.\ in rigorous test matter approximation, and 
the apparent violation of WCCC arises because effects beyond the lowest order 
are included in some way, 
but only partially. The restoration of cosmic censorship is 
achieved by properly taking into account all relevant effects.  
For example, in \cite{Hod:2008zza} the overspinning of a near extremal Reissner--Nordstr\"om black hole by waves carrying angular momentum was considered. Such a setting immediately implies the inclusion of effects beyond lowest order,
because  $\eta$ depends on $J$ through $J^2$, thus in the lowest order the $\eta$ parameter of 
a Reissner--Nordstr\"om black hole cannot be changed by changing its angular momentum. In \cite{Hod:2008zza} it was shown that although an apparent violation of the WCCC can be found if the change of $\eta$ due to the change of $J$ is not neglected but the waves are assumed to propagate on fixed Reissner--Nordstr\"om background, cosmic censorship 
is restored if the effect of the waves on the background during the interaction process is also taken into account, as required by the consistency of the approximation applied.  
In \cite{JS} the overspinning of a slightly subextremal Kerr black hole with a test body was considered, and also in this study some higher order quantities were not neglected, while radiative and self-force effects were not taken into account.
Later in \cite{BCK1,BCK2,CB,Colleoni:2015ena} it was argued that self-force effects are not negligible in this scenario and they might be the main effect preventing the violation of the WCCC.

In \cite{Hod:2013vj} a further interesting effect is described; the formation of another horizon outside
a Reissner--Nordstr\"om black hole when a charged shell that would be expected to destroy it 
is adiabatically lowered towards its event horizon. 
This scenario is hard to compare with the case of the Dirac field, but even if a similar effect can show up also in the 
latter case, it is a higher order effect, therefore it cannot be expected to completely override the lowest order result. Regarding the adiabatic lowering of charged objects, 
it should also be noted that it is not necessary to assume that 
the particle comes from infinity
in the derivations in \cite{Wald2,Needham, TG} 
of the result that in rigorous test particle approximation 
the cosmic censorship principle is respected.

\section{Superradiance}
\label{sec.srad}

The setting in which the phenomenon of black hole superradiance 
occurs is similar to that of the thought experiment, with the difference
that the initial black hole is not necessarily extremal and 
the quantity of interest is the total energy $dE$
that flows through the event horizon, instead of $d\eta$. 
Superradiance occurs if $dE$ has a sign 
that corresponds to an amplification of the energy 
of the field outside the event horizon. 
In addition, the angular momentum and the electric charge of the field 
can also be considered in the study of superradiance. 

In the literature it is usual to 
describe superradiance in terms of the amplitude of suitable radial functions 
which arise when the complete separation of the variables 
is carried out (see e.g.\ \cite{Ch,sr,RWPU}), 
but in this section we do not use these amplitudes, 
in accordance with our aim to avoid the use of the 
complete separation of variables as much as possible.    

The superradiance of individual energy and angular momentum modes 
can also be defined. In this case the quantity that determines
if a certain mode is superradiant
is the sign of the rate $\frac{dE}{d\tau}$.

\subsection{Absence of superradiance of Dirac fields}
\label{sec.srad1}

If some matter described by the Dirac field is thrown into a black hole, then the total
electric charge absorbed by the black hole is  
\beq
dQ  =  \int_{-\infty}^\infty d\tau \int_H \sqrt{-g}\, e j^r\, d\theta d\varphi\ .
\eeq
By definition $j^r = \zeta_\mu j^\mu$, and one can argue, in the same way as in Section \ref{sec.ft},
that at the event horizon $\zeta^\mu$ is past directed and null, 
$j^\mu$ is always future directed and null or time-like, thus $j^r\le 0$, and so 
$e\, dQ <0$. This means that the total electric charge 
falling through the event horizon always has the same sign as 
the charge of the Dirac field, 
thus the charge outside the event horizon does not increase. In other words,
the Dirac field does not show superradiance in relation to electric charge.

Considering energy and angular momentum, using the Fourier expansion (\ref{eq.fourier})
one finds that 
the total energy and angular momentum absorbed by the 
black hole is
\bea
dE & = & -\int_{-\infty}^\infty d\tau \int_H \sqrt{-g}\, \mc{E}^r\, d\theta d\varphi \nonumber\\
\label{eq.63}
& = & (2\pi)^2  \sum_n \int d\omega \int_H d\theta\, \sqrt{-g}\  (-\omega\, \bar{\psi}_{\omega,n} \gamma^r \psi_{\omega,n})\\
\label{eq.64}
dL & = & \phantom{-}\int_{-\infty}^\infty d\tau \int_H \sqrt{-g}\, \mc{J}^r\, d\theta d\varphi \nonumber \\
& = & (2\pi)^2  \sum_n \int d\omega \int_H d\theta\, \sqrt{-g}\  (-n\, \bar{\psi}_{\omega,n} \gamma^r \psi_{\omega,n})\ .
\eea
In the derivation of (\ref{eq.63}) and (\ref{eq.64}) it is useful to write (\ref{eq.energy}) and (\ref{eq.jmod}) as 
$\mc{E}^\mu = 
\frac{1}{2}(\bar{\Psi} \ii \gamma^\mu \partial_t \Psi -\partial_t\bar{\Psi}\ii \gamma^\mu\Psi)$
and
$\mc{J}^\mu = 
\frac{1}{2}(\bar{\Psi} \ii \gamma^\mu \partial_\varphi \Psi -\partial_\varphi\bar{\Psi}\ii \gamma^\mu\Psi)
-eQ_mCj^\mu$, because the vector potential does not appear explicitly in the latter expressions.

Since, as explained in Section \ref{sec.ft},  
$\bar{\psi}_{\omega,n} \gamma^r \psi_{\omega,n}\le 0$ at the event horizon,
the integrands in (\ref{eq.63}) and (\ref{eq.64}) have 
the same signs as $\omega$ and $n$, respectively. 
Consequently, if the Fourier expansion of $\Psi$ contains 
only positive frequency modes, 
then the energy falling through the event horizon 
is also positive and the energy outside the event horizon 
does not increase, and analogous statements can be made 
for negative frequency modes and for angular momentum.  
This means that the Dirac field is not superradiant 
in relation to energy and angular momentum either. 

Instead of considering solutions of the form (\ref{eq.fourier}), one can consider waves
consisting of a single mode, $\ee^{-\ii\omega \tau}\ee^{\ii (n- C eQ_m)\varphi}\psi_{\omega,n}(r,\theta)$. 
$dQ$, $dE$ and $dL$ are not meaningful for such waves, but one can study the rates
$\frac{dQ}{d\tau} = \int_H \sqrt{-g}\, e j^r\, d\theta d\varphi $, 
$\frac{dE}{d\tau} = -\int_H \sqrt{-g}\, \mc{E}^r\, d\theta d\varphi$, 
$\frac{dL}{d\tau} =  \int_H \sqrt{-g}\, \mc{J}^r\, d\theta d\varphi$.  
These rates can be expressed as
\bea
\label{eq.62b}
\frac{dQ}{d\tau}  & = & 2\pi  \int_H d\theta\, \sqrt{-g}\  (e\, \bar{\psi}_{\omega,n} \gamma^r \psi_{\omega,n})\\
\label{eq.63b}
\frac{dE}{d\tau} & = & 2\pi  \int_H d\theta\, \sqrt{-g}\  (-\omega\, \bar{\psi}_{\omega,n} \gamma^r \psi_{\omega,n})\\
\label{eq.64b}
\frac{dL}{d\tau} & = & 2\pi \int_H d\theta\, \sqrt{-g}\  (-n\, \bar{\psi}_{\omega,n} \gamma^r \psi_{\omega,n})\ .
\eea
Using these expressions one can argue
in the same way as above 
that the Dirac field does not have superradiant modes.

\subsection{Superradiant frequency range of scalar fields}
\label{sec.srad2}

The massive complex scalar field has the Lagrangian density
\beq
\label{eq.lagrsc}
\mc{L}=g^{\mu\nu}(\partial_\mu-\ii eA_\mu)\Phi^*(\partial_\nu+\ii eA_\nu)\Phi-m^2\Phi^*\Phi
\eeq
and the corresponding field equation
$(\nabla^\mu +\ii e A^\mu)  (\nabla_\mu+\ii e A_\mu) \Phi = -m^2 \Phi$.
The electric current of the scalar field is
\beq
j^\mu= \frac{\partial \mc{L}}{\partial A_\mu} =  
-\ii e[\Phi^*(\partial^\mu+\ii eA^\mu)\Phi-\Phi(\partial^\mu-\ii eA^\mu)\Phi^*]\ .
\eeq
In \cite{TG} we found the
energy and angular momentum current densities 
$\mc{E}^\mu$ and $\mc{J}^\mu$ by applying Noether's theorem.
The result for $\mc{J}^\mu$ had to be modified in the same way as 
in Section \ref{sec.ft} to eliminate its dependence on the 
gauge parameter $C$. 
$\mc{E}^\mu$ and $\mc{J}^\mu$ are given by the expressions
\beq
\mc{E}^\mu =  \hat{T}\indices{^\mu_\tau} - A_\tau j^\mu\ ,\qquad
\mc{J}^\mu =  \hat{T}\indices{^\mu_\varphi} - (A_\varphi - Q_m C)j^\mu\ ,
\eeq
where
\beq
\label{eq.tmnsc}
\hat{T}_{\mu\nu}=
(\partial_\mu-\ii eA_\mu)\Phi^* (\partial_\nu +\ii eA_\nu)\Phi
+ (\partial_\mu+\ii eA_\mu)\Phi (\partial_\nu -\ii eA_\nu)\Phi^* - g_{\mu\nu}\mc{L}\ .
\eeq
Using the Fourier expansion
\beq
\label{eq.fourierphi}
\Phi= \sum_n\, \int d\omega\ \ee^{-\ii\omega \tau}\ee^{\ii (n- C eQ_m)\varphi}\phi_{\omega,n}(r,\theta)
\eeq
of $\Phi$,
where $\ee^{-\ii\omega \tau}\ee^{\ii (n- C eQ_m)\varphi}\phi_{\omega,n}(r,\theta)$ 
are solutions of the field equation,
one obtains the following results for $dQ$, $dE$ and $dL$:
\bea
dQ & = &  -\int_{-\infty}^\infty d\tau \int_H \sqrt{-g}\, j^r\, d\theta d\varphi \nonumber \\
& = &  -(2\pi)^2 \sum_n\, \int d\omega\ \int_H d\theta\  2 (a^2+r_+^2) \sin\theta\, \phi_{\omega,n}^* \phi_{\omega,n}\, e \tilde{\omega} 
\label{eq.z1}\\
dE & = & -\int_{-\infty}^\infty d\tau \int_H \sqrt{-g}\, \mc{E}^r\, d\theta d\varphi \nonumber \\ 
& = & (2\pi)^2 \sum_n\, \int d\omega\ \int_H d\theta\  2 (a^2+r_+^2) \sin\theta\, \phi_{\omega,n}^* \phi_{\omega,n}\, \omega \tilde{\omega}
\label{eq.z2}\\
dL & = & \phantom{-}\int_{-\infty}^\infty d\tau \int_H \sqrt{-g}\, \mc{J}^r\, d\theta d\varphi \nonumber \\ 
& = & (2\pi)^2 \sum_n\, \int d\omega\ \int_H d\theta\  2 (a^2+r_+^2) \sin\theta\, \phi_{\omega,n}^* \phi_{\omega,n}\, n \tilde{\omega}\ ,
\label{eq.z3}
\eea
where $\tilde{\omega}$ is defined as in (\ref{eq.to}).
In the derivation of (\ref{eq.z1}), (\ref{eq.z2}) and (\ref{eq.z3}) only 
the derivatives $\partial_\tau\Phi$ and $\partial_\varphi\Phi$ of $\Phi$ appear, 
which can be evaluated easily,
and one can also use (\ref{eq.chi})-(\ref{eq.rho}) and (\ref{eq.sqrtg}). 
Due to the factor $g_{\mu\nu}$, the $-g_{\mu\nu}\mc{L}$ term appearing 
in (\ref{eq.tmnsc}) does not give any contribution 
to $dE$ and $dL$, as $\delta\indices{^r_\tau}=\delta\indices{^r_\varphi}=0$.
For 
$\ee^{-\ii\omega \tau}\ee^{\ii (n- C eQ_m)\varphi}\phi_{\omega,n}(r,\theta)$ 
to be single valued for both $C=1$ and $C=-1$,
both $n- eQ_m$ and $n+ eQ_m$ have to be integer, 
implying that $n$ and $eQ_m$ are either integer or half-integer, 
thus in the summations $n$ should take integer values if $eQ_m$ is integer 
and half-integer values if $eQ_m$ is half-integer \cite{SemizKG}.
It should also be noted that far from the black hole only modes with $|\omega|>m$ describe propagating waves.

Since $\phi_{\omega,n}^* \phi_{\omega,n}$ is positive 
unless $\phi_{\omega,n}=0$, (\ref{eq.z2}) shows that $dE$ is negative 
if the main contribution to the integral comes from the 
frequency range where 
\beq
\label{eq.srad}
\omega \tilde{\omega}<0\ .
\eeq
In this case the scalar field exhibits superradiance 
in the sense that the total energy of the 
field outside the event horizon increases. 
The sign of the integrands in (\ref{eq.z1}) and (\ref{eq.z3}) 
is also determined by $e\tilde{\omega}$ and $n\tilde{\omega}$, respectively, 
instead of $e\omega$ and $n\omega$.

For individual modes $\ee^{-\ii\omega \tau}\ee^{\ii (n- C eQ_m)\varphi}\phi_{\omega,n}(r,\theta)$, 
the rates $\frac{dE}{d\tau}$, $\frac{dL}{d\tau}$ and $\frac{dQ}{d\tau}$ can be expressed as
\bea
\frac{dQ}{d\tau} & = & -2\pi  \int_H d\theta\  2 (a^2+r_+^2) \sin\theta\, \phi_{\omega,n}^* \phi_{\omega,n}\, e \tilde{\omega} 
\label{eq.z1b}\\
\frac{dE}{d\tau} & = & \phantom{-}2\pi  \int_H d\theta\  2 (a^2+r_+^2) \sin\theta\, \phi_{\omega,n}^* \phi_{\omega,n}\, \omega \tilde{\omega}
\label{eq.z2b}\\
\frac{dL}{d\tau} & = & \phantom{-}2\pi  \int_H d\theta\  2 (a^2+r_+^2) \sin\theta\, \phi_{\omega,n}^* \phi_{\omega,n}\, n \tilde{\omega}\ .
\label{eq.z3b}
\eea
Of course, the superradiant frequencies following from (\ref{eq.z2b}) 
are the same as those that follow from (\ref{eq.z2}).

\section{Conclusion}
\label{sec.concl}

We showed that a dyonic Kerr--Newman black hole
can be converted into a naked singularity by
interaction with massive charged classical Dirac fields, generalizing a recent result
\cite{Duztas1} which applies to massless Dirac fields and Kerr black holes.  
We found that for this conversion the spectrum of the Dirac field has to be dominated
by modes with temporal and angular frequencies $\omega$ and $n$ satisfying the inequality 
(\ref{eq.to}).
We also showed that the 
creation of a naked singularity is possible because the  
null energy condition is not satisfied by the Einstein--Hilbert energy-momentum
tensor of the Dirac field. 
This means that the classical Dirac field 
does not satisfy the criteria under which 
the WCCC is expected to hold, thus in strict sense the possibility 
found in \cite{Duztas1} and in the present paper 
does not contradict the WCCC. 
These features of the Dirac field are complementary to those of 
the scalar and the electromagnetic field, 
which satisfy the null energy condition and as a consequence 
cannot convert a black hole into a naked singularity.

We gave a derivation of the absence of superradiance of the Dirac field 
around dyonic Kerr--Newman black holes,
and we determined the temporal and angular frequencies 
for which the scalar field is superradiant.
We found that the superradiant modes are those that satisfy the condition (\ref{eq.srad}). 
The frequencies satisfying (\ref{eq.to}) partially agree with those satisfying (\ref{eq.srad}).

Although the well-known separability of the scalar wave equation and of the 
Dirac equation around black holes is very useful 
for several purposes and is often used also in the 
discussion of the testing of the WCCC or of the superradiance phenomenon, 
its use could be largely avoided in this paper. Nevertheless, as we used it
in an argument at the end of Section \ref{sec.ft}, we described it  
in an appendix in horizon-penetrating coordinates.

In the derivations the test field approximation was applied, but we argued that 
the destruction of black holes cannot be completely prevented by backreaction effects. 
It was also assumed that the final state that arises after the interaction of the 
black hole and the Dirac field is again a dyonic Kerr--Newman configuration, 
without any Dirac hair. This assumption is in accordance with the no-hair conjecture, 
but its validity would nevertheless be interesting to investigate.

It is natural to hope that the formation of a naked singularity can be ruled out 
in quantum theory, since problems related to negative 
energies are absent in quantum field theory in Minkowski spacetime. 
Results on quantum effects
have already appeared in the literature \cite{Unruh,RS2,RS,Hod2,Duztas:2015qqa}, 
and comments can be found also in \cite{Duztas1}. 
We leave further investigations of the quantum Dirac field for future work.

\section*{Acknowledgments}

The author is supported by an MTA Lend\"ulet grant. This work was also supported in part by the grant
Die Aktion \"Osterreich-Ungarn, Wissenschafts- und Erziehungskooperation 90\"ou1.

\appendix

\renewcommand{\theequation}{\Alph{section}.\arabic{equation}} 
\setcounter{equation}{0}

\section{Dirac spinor fields in curved spacetime}
\label{sec.A}

In this paper we apply the tetrad formalism to incorporate Dirac spinor fields into general relativity,
as described e.g.\ in \cite{W}.

We denote internal Lorentz vector indices by letters with an overbar.
In contrast with Lorentz and spacetime vector indices, 
the normal position of Dirac spinor indices is taken to be the lower one, 
and hence cospinor indices are in the upper position.
Dirac spinor indices are often not written out explicitly. 

Lorentz vector and spinor indices are scalar indices 
with respect to spacetime diffeomorphisms;
in particular, tensors having only Lorentz vector and 
spinor indices are scalars under spacetime diffeomorphisms.
On the other hand, local Lorentz transformations do not act on the spacetime manifold 
and spacetime vector indices are scalar indices with respect to them.

We use orthonormal tetrad fields $V^{\bar{\mu}}_\mu$, i.e.\
$g^{\bar{\mu}\bar{\nu}}=V^{\bar{\mu}}_\mu V^{\bar{\nu}}_\nu g^{\mu\nu}$,
$g^{\bar{\mu}\bar{\nu}} = \mathrm{diag}(1,-1,-1,\\ -1)$.

The Levi--Civita covariant differentiation is extended to 
tensors with Lorentz vector indices in the following way:
\beq
\nabla_\mu v^{\bar{\nu}}  =  \partial_\mu v^{\bar{\nu}}+\mc{S}\indices{^{\bar{\nu}}_{\bar{\lambda}\mu}}v^{\bar{\lambda}}\ ,
\qquad
\mc{S}\indices{^{\bar{\lambda}}_{\bar{\eta}\mu}}=-V_{\bar{\eta}}^\nu\nabla_\mu^{LC}V_\nu^{\bar{\lambda}}\ ,
\eeq
where $\mc{S}\indices{^{\bar{\lambda}}_{\bar{\eta}\mu}}$ 
is an analogue of the Christoffel symbols.
On the right hand side the superscript $LC$ indicates that 
the Levi--Civita covariant differentiation should be used.
With this definition the covariant derivative of the Lorentz metric tensor 
and of the tetrad field is zero:
$\nabla_\mu g_{\bar{\nu}\bar{\lambda}}=0$, $\nabla_\mu V_{\nu}^{\bar{\lambda}}=0$.

The Dirac gamma matrices $\gamma^\mu$ are defined as
\beq
\gamma^\mu=V^\mu_{\bar{\mu}}\gamma^{\bar{\mu}}\ ,
\eeq
where $\gamma^{\bar{\mu}}$ are the standard Minkowski space gamma matrices. 
We use the Weyl representation for them, 
\beq
\gamma^{\bar{0}}=\left( \begin{array}{cc}
0 & I \\
I & 0 \\
\end{array}\right),\qquad
\gamma^{\bar{i}}=\left( \begin{array}{cc}
0 & \sigma^i \\
-\sigma^i & 0 \\
\end{array}\right),\qquad i=1,2,3\ ,
\eeq
where $I$ denotes the $2\times 2$ identity matrix and 
$\sigma^i$ are the Pauli sigma matrices (see \cite{PS}). 
$\gamma^\mu$ has the property
$\{\gamma^{\mu},\gamma^{\nu} \}=2g^{\mu\nu}$,
where $\{\ ,\, \}$ denotes the anticommutator $\{A,B\} = AB+BA$.

The Levi--Civita connection can be extended to tensors having Dirac spinor indices 
in the following way:
\beq
\nabla_\mu \psi_\alpha  =  \partial_\mu \psi_\alpha   + S\indices{_\alpha^\beta_\mu} \psi_\beta\ ,
\eeq
where
\beq
S_\mu = \frac{1}{4}\sigma^{\bar{\nu}\bar{\lambda}}\mc{S}_{\bar{\nu}\bar{\lambda}\mu}\ ,\qquad
\sigma^{\bar{\mu}\bar{\nu}}=\frac{1}{2}[\gamma^{\bar{\mu}},\gamma^{\bar{\nu}}]\ .
\eeq
With this definition of the covariant differentiation of spinors the covariant derivative of the 
gamma tensor is zero:
$\nabla_\mu \gamma^\nu =0$.

In the Weyl representation the Dirac conjugation for Dirac spinors with a lower spinor index 
is customarily expressed as
$\bar{\psi}=\psi^{*T} \gamma^{\bar{0}}$,
where the ${}^T$ denotes transposition, 
the ${}^*$ denotes componentwise complex conjugation, and as usual 
the Dirac conjugation is denoted by an overbar. 

Finally we note that 
we do not introduce any raising and lowering convention for spinor indices.

\section{Separation of variables for the Dirac equation}
\label{sec.B}

In this appendix the separation of variables 
for the Dirac equation in dyonic Kerr--Newman background is described.
This is done in the horizon-penetrating coordinates $(\tau,r,\theta,\varphi)$, 
using the Kinnersley-type tetrad $\tilde{V}_\mu^{\bar{\mu}}$ introduced in 
Section \ref{sec.tetrad} and the Weyl representation 
for the Dirac gamma matrices described in \ref{sec.A}.   
The separability of the Dirac equation in dyonic Kerr--Newman background 
was shown first in \cite{Semiz3}, 
in Boyer--Lindquist coordinates. 
The reader is referred to this article and to \cite{sr,DD} 
for further references to earlier results.
The recent article \cite{Roken:2015fja} focuses on the separability of 
the Dirac equation in horizon-penetrating coordinates,  
but only in Kerr geometry. 

The asymptotic behaviour at the event horizon of the solutions of the 
radial equations that arise after the separation of variables
is also discussed in a subsection.

The application of the method of separation of variables 
to finding solutions of the Dirac equation
in dyonic Kerr--Newman background begins with 
assuming that $\Psi$ depends on $\tau$ and $\varphi$ harmonically as
\beq
\Psi(\tau,r,\theta,\varphi)= \ee^{-\ii\omega \tau}\ee^{\ii (n- C eQ_m)\varphi}\psi(r,\theta)\ ,
\eeq
where $\omega$ and $n$ denote the temporal and angular frequency, respectively. 
The term  $-CeQ_m$ in the factor $\ee^{\ii (n- C e Q_m)\varphi}$ 
is included because of the gauge transformation
done at equation (\ref{eq.am2}) in Section \ref{sec.hp} (see also \cite{Semiz3}).
After introducing the new field variables $f_i,\, i=1,\dots, 4$, as
\beq
\label{eq.f}
f_1=\psi_1,\ \ \ f_2=\frac{1}{r}(r-\ii a \cos\theta)\psi_2,\ \ \ f_3=\frac{1}{r}(r+\ii a \cos\theta)\psi_3,\ \ \ f_4=\psi_4,
\eeq
where $\psi_i,\, i=1,\dots, 4$, denote the components of $\psi$, 
the Dirac equation can be written in the form 
\bea
\label{eq.d1}
D_+ f_3 + L_+ f_4 & = & -\ii m (r+\ii a \cos\theta) f_1\\
D_- f_4 + L_- f_3 & = & -\ii m  (r+\ii a \cos\theta) f_2\\
D_- f_1 - L_+ f_2 & = &  -\ii m (r-\ii a \cos\theta) f_3\\
D_+ f_2 - L_- f_1 & = & -\ii m (r-\ii a \cos\theta) f_4\ ,
\label{eq.d4}
\eea
where
\bea
D_+ & = &   \frac{\sqrt{2}}{r^2}[-Mr +r^2 + \ii(2anr-2eQ_er^2 - 2a^2r\omega -2r^3\omega +r\Delta\omega) \nonumber\\
&& +r\Delta \partial_r]\\ 
D_- & = & \frac{-1}{\sqrt{2}}(1+\ii r\omega +r\partial_r)\\
L_+ & = & \frac{1}{2\sin\theta}[-2n + (1+2eQ_m)\cos\theta +a\omega(1-\cos 2\theta)] +\partial_\theta \\
L_- & = & \frac{1}{2\sin\theta}[2n + (1-2eQ_m)\cos\theta -a\omega(1-\cos 2\theta)] +\partial_\theta\ .
\eea
We note that in (\ref{eq.f}) the factors in front of $\psi_2$ and $\psi_3$ 
are chosen so that in the $r\to\infty$ limit 
$f_2\to\psi_2$ and $f_3\to\psi_3$.

By taking the ansatz 
\bea
\label{eq.ans1}
&&f_1(r,\theta)=R_+(r) S_+(\theta)\qquad f_2(r,\theta)=R_-(r) S_-(\theta)\\
\label{eq.ans2}
&&f_3(r,\theta)= R_-(r) S_+(\theta)\qquad f_4(r,\theta)= R_+(r) S_-(\theta)
\eea
the $r$ and $\theta$ parts of equations (\ref{eq.d1})-(\ref{eq.d4}) 
become separated and one finds the ordinary differential equations
\bea
\label{eq.rr1}
D_+ R_- -(\lambda -\ii m r )R_+ & = & 0\\
\label{eq.rr2}
D_- R_+ +(\lambda +\ii m r )R_- & = & 0\\
\label{eq.ll1}
L_+ S_- +(\lambda - m a \cos\theta ) S_+ & = & 0\\
\label{eq.ll2}
L_- S_+ -(\lambda + m a \cos\theta ) S_- & = & 0
\eea
for $R_+$, $R_-$, $S_+$ and $S_-$,
where $\lambda$ is the separation constant. 
Initially one introduces different separation constants in each equation
(\ref{eq.d1})-(\ref{eq.d4}), but then one sees that they have to be related. 

By eliminating $R_+$ or $R_-$ and $S_+$ or $S_-$ one gets the 
second order decoupled equations
\bea
\label{eq.rad1}
D_- D_+ R_- - \frac{\ii m r}{\sqrt{2}(\lambda-\ii m r)}D_+ R_- +(\lambda^2 + m^2 r^2) R_- & = & 0\\
\label{eq.rad2}
D_+ D_- R_+ - \frac{\ii \sqrt{2}\, m \Delta}{r(\lambda + \ii m r)}D_- R_+ + (\lambda^2 + m^2 r^2) R_+ & = & 0
\eea
and
\bea
\label{eq.l1}
L_-L_+ S_-  - \frac{ma\sin\theta}{\lambda-ma\cos\theta }L_+ S_- +(\lambda^2-m^2a^2\cos^2\theta) S_-   & = & 0\\
L_+L_- S_+  + \frac{ma\sin\theta}{\lambda+ma\cos\theta }L_- S_+ +(\lambda^2-m^2a^2\cos^2\theta) S_+  & = & 0\ .
\label{eq.l2}
\eea
It is useful to write out the explicit form of $L_- L_+$,  $L_+ L_-$, $D_- D_+$, $D_+ D_-$:
\bea
L_- L_+ S_- & = & -\frac{1}{8\sin^2\theta}\Big[3+8n^2+8eQ_m+4e^2Q_m^2-8an\omega+3a^2\omega^2 \nonumber\\
&& -2(n(4+8eQ_m)+a(1-2eQ_m)\omega)\cos\theta  \nonumber\\
&& +(-1+4e^2Q_m^2+8an\omega-4a^2\omega^2)\cos 2\theta \nonumber\\
&&+(2a\omega-4aeQ_m\omega)\cos 3\theta +a^2\omega^2\cos 4\theta\Big]S_- \nonumber\\
&&+\frac{\cos\theta}{\sin\theta}\partial_\theta S_- + \partial_\theta^2 S_-\ ,
\eea
\bea
L_+ L_- S_+ & = & -\frac{1}{8\sin^2\theta}\Big[3+8n^2-8eQ_m+4e^2Q_m^2-8an\omega+3a^2\omega^2 \nonumber\\
&& +2(n(4-8eQ_m)+a(1+2eQ_m)\omega)\cos\theta  \nonumber\\
&& +(-1+4e^2Q_m^2+8an\omega-4a^2\omega^2)\cos 2\theta \nonumber\\
&&-(2a\omega+4aeQ_m\omega)\cos 3\theta +a^2\omega^2\cos 4\theta\Big]S_+ \nonumber\\
&&+\frac{\cos\theta}{\sin\theta}\partial_\theta S_+ + \partial_\theta^2 S_+\ ,
\eea
and 
\bea
D_+ D_- R_+ & = & \Big[-1+2\ii e Q_e +\frac{M-2\ii an}{r}+\omega(\ii M+\ii r+2an-2eQ_e r)\nonumber\\
&&+\frac{2\ii \omega(a^2-\Delta)}{r}
+\omega^2(\Delta-2a^2-2r^2)\Big] R_+ \nonumber\\
&&+\Big[M-r-\frac{2\Delta}{r}+2\ii(-an+e Q_e r)+2\ii\omega(a^2+r^2-\Delta)\Big]\partial_r R_+ \nonumber \\
&& -\Delta \partial_r^2 R_+\ ,
\eea
\bea
D_- D_+ R_- & = & \big[-1+2\ii e Q_e +\omega(3\ii M +\ii r+2an-2eQ_e r)\nonumber\\
&&
+\omega^2(\Delta-2a^2-2r^2)\big] R_- \nonumber\\
&&+\big[3(M-r)+2\ii(-an+e Q_e r)+2\ii\omega(a^2+r^2-\Delta)\big]\partial_r R_- \nonumber\\
&& -\Delta \partial_r^2 R_-\ .
\eea

If the mass of the Dirac field is zero, then after
multiplying by the integrating factor $\sin\theta$ 
the  angular equations (\ref{eq.l1}) and (\ref{eq.l2}) 
take a Sturm--Liouville form.
The weight factor appearing in the scalar product 
for the solutions is also $\sin\theta$.
Although the case $m\ne 0$ is more complicated, 
it was studied e.g.\ in \cite{SFC,KM,Finster:1999ry,Finster:2000jz,DG} at $Q_m=0$.

In the $r\to\infty$ limit the radial equations (\ref{eq.rad1}) and (\ref{eq.rad2}) become 
\beq
\label{eq.pw}
-\omega^2 R_\pm -\partial_r^2 R_\pm +m^2 R_\pm =0\ .
\eeq

\subsection{Asymptotic behaviour of the radial functions at the\\ event horizon}
\label{sec.B.1}

The second order radial equations (\ref{eq.rad1}) and (\ref{eq.rad2}) 
are singular at the event horizon; 
the singularity is regular if the black hole is not extremal 
and irregular if the black hole is extremal. 
By analyzing the asymptotic behaviour of the solutions 
near the event horizon one finds two kinds of asymptotic behaviour.
One of them is such that $R_+$ (or $R_-$) 
approaches a finite nonzero value as $r\to r_+$, the other one is such that
$R_+$ goes to zero and $|R_-|$ to infinity as $r\to r_+$. Solutions with the latter
behaviour can be considered unphysical.

More specifically, in the non-extremal case $R_+$ takes the form 
\beq
c_1 (r-r_+)^{s_1} y_1(r-r_+) + c_2 (r-r_+)^{s_2} y_2(r-r_+)\ ,
\eeq 
where $c_1$ and $c_2$ are integration constants, 
$y_1$ and $y_2$ are functions that are regular and nonzero at $0$, 
and the characteristic exponents $s_1$ and $s_2$ are solutions of the indicial equation
\beq
s(s-1)-\frac{1}{r_+-r_-}[M-r_+ +2\ii (a^2+r_+^2) \tilde{\omega}]s=0\ .
\eeq
The solutions of this equation are
\beq
\label{eq.chexp1}
s_1=0,\qquad s_2=\frac{1}{2}+\ii \frac{2(a^2+r_+^2)\tilde{\omega}}{r_+-r_-}\ ,
\eeq
thus the solution corresponding to $c_2=0$ is regular and nonzero at the event horizon, 
whereas the solution corresponding to $c_1=0$ is not regular but is vanishing at $r_+$. 
The latter solution can also be written as
\beq
\ee^{s_2\log(r-r_+)} y_2(r-r_+)\ ,
\eeq
showing that near the event horizon the absolute value of this solution behaves like
$\sim(r-r_+)^{1/2}$ and the oscillation frequency of its phase increases to infinity as $r\to r_+$.   

In the extremal case the asymptotic behaviour of the 
$R_+\to 0$ type solution of (\ref{eq.rad2}) at the event horizon
is found to be
\beq
\sim\exp\left[-\frac{2\ii(a^2+r_+^2)\tilde{\omega}}{r-r_+} +(1+2\ii e Q_e + 4\ii\omega r_+)\log(r-r_+)\right]\ ,
\eeq
showing that the absolute value of this solution behaves like $\sim(r-r_+)$ 
and the oscillation frequency of its
phase increases to infinity as $r\to r_+$.

The characteristic exponents for $R_-$ in the non-extremal case are
\beq
s_1=0,\qquad s_2=-\frac{1}{2}+\ii \frac{2(a^2+r_+^2)\tilde{\omega}}{r_+-r_-}\ ,
\eeq
as can be expected from (\ref{eq.rr2}) and (\ref{eq.chexp1}).
Similarly, in the extremal case the asymptotic behaviour of the 
$|R_-|\to \infty$ type solution of (\ref{eq.rad1}) 
at the event horizon is
\beq
\sim\exp\left[-\frac{2\ii(a^2+r_+^2)\tilde{\omega}}{r-r_+} +(-1+2\ii e Q_e + 4\ii\omega r_+)\log(r-r_+)\right]\ .
\eeq


\small
\begin{thebibliography}{99}


\bibitem{Penrose}Penrose R 1969 
Gravitational collapse: the role of general relativity 
\emph{Riv. Nuovo Cimento} \textbf{1}, special number, 252


\bibitem{WaldGR}Wald R M 1984  
\textsl{General relativity} 
(Chicago: University of Chicago Press)


\bibitem{Wald0}Wald R M 1997
Gravitational collapse and cosmic censorship, 
arXiv: gr-qc/9710068

\bibitem{Joshi}Joshi P S 2002
Cosmic censorship: a current perspective 
\emph{Modern Phys. Lett.} A \textbf{17} 1067,
arXiv: gr-qc/0206087

\bibitem{Clarke}Clarke C J S 1994
A title of cosmic censorship
\emph{Class. Quantum Grav.} \textbf{11} 1375

\bibitem{Singh}Singh T P 1999 
Gravitational collapse, black holes and naked singularities
\emph{J. Astrophys. Astron.} \textbf{20} 221, 
arXiv: gr-qc/9805066

\bibitem{Krolak}Krolak A 1999
Nature of singularities in gravitational collapse
\emph{Prog. Theor. Phys. Suppl.} \textbf{136} 45


\bibitem{Wald2}Wald R M 1974 
Gedanken experiments to destroy a black hole
\emph{Annals of Physics} \textbf{83} 548



\bibitem{Needham}Needham T 1980 
Cosmic censorship and test particles
\emph{Phys. Rev.} D \textbf{22} 791 


\bibitem{Hiscock}Hiscock W A 1981  
Magnetic charge, black holes and cosmic censorship 
\emph{Ann. Phys.} \textbf{131} 245


\bibitem{Semiz2}Semiz I 1990 
Dyon black holes do not violate cosmic censorship
\emph{Class. Quantum Grav.} \textbf{7} 353


\bibitem{Semiz1}Semiz I 2011 
Dyonic Kerr--Newman black holes, complex scalar field and cosmic censorship
\emph{Gen. Rel. Grav.} \textbf{43} 833,
arXiv: gr-qc/0508011

\bibitem{Duztas2}Duztas K 2014 
Electromagnetic field and cosmic censorship
\emph{Gen. Rel. Grav.} {\bf 46} 1709,
arXiv:1312.7361 [gr-qc]

\bibitem{DS}Duztas K and Semiz I 2013 
Cosmic censorship, black holes and integer-spin test fields
\emph{Phys. Rev.} D {\bf 88} 064043,
arXiv:1307.1481 [gr-qc]

\bibitem{Duztas1}Duztas K 2015 
Stability of event horizons against neutrino flux: the classical picture
\emph{Class. Quantum Grav.} {\bf 32} 075003,
arXiv:1408.1735 [gr-qc]

\bibitem{TG}Toth G Z 2012 
Test of the weak cosmic censorship conjecture with a charged scalar field and dyonic Kerr--Newman black holes
\emph{Gen. Rel. Grav.} {\bf 44} 2019,
arXiv:1112.2382 [gr-qc]

\bibitem{Ch}Chandrasekhar S 1983
\emph{The Mathematical Theory of Black Holes} (New York: Oxford University Press) 

\bibitem{sr}Brito R, Cardoso V and Pani P 2015 
Superradiance
\emph{Lect. Notes Phys.} {\bf 906}, 
arXiv:1501.06570 [gr-qc]

\bibitem{DD}Dolan S R and Dempsey D 2015 
Bound states of the Dirac equation on Kerr spacetime
\emph{Class. Quantum Grav.} {\bf 32} 18  184001,
arXiv:1504.03190v1 [gr-qc] 

\bibitem{Lee}Lee C H 1977 
Massive spin-$\frac{1}{2}$ wave around a Kerr--Newman black hole
\emph{Phys. Lett.} B {\bf 68} 152

\bibitem{Unruh:1973}Unruh W 1973
Separability of the Neutrino Equations in a Kerr Background
\emph{Phys. Rev. Lett.}  {\bf 31} 20,  1265

\bibitem{Gueven:1977}Gueven R 1977
Wave Mechanics of Electrons in Kerr Geometry
\emph{Phys. Rev.} D {\bf 16} 1706
 

\bibitem{Casals:2012es}
Casals M, Dolan S R, Nolan B C, Ottewill A C and Winstanley E 2013
Quantization of fermions on Kerr space-time
\emph{Phys. Rev.} D {\bf 87} 6  064027,
arXiv:1207.7089 [gr-qc]

\bibitem{Arderucio:2014oua}
Arderucio B,
Superradiance: Classical, Relativistic and Quantum Aspects, 
arXiv:1404.3421 [gr-qc], Section 3.4
 


\bibitem{SFC}Suffern K G, Fackerell E D and Cosgrove C M 1983 
Eigenvalues of the Chandrasekhar--Page angular functions 
\emph{J. Math. Phys.} {\bf 24} 1350



\bibitem{KM}Kalnins E G and Miller W Jr. 1992 
Series solutions for the Dirac equation in Kerr--Newman space-time 
\emph{J. Math. Phys.} {\bf 33} 286



\bibitem{Finster:1999ry}
Finster F, Kamran N, Smoller J and Yau S T 2000
Nonexistence of time periodic solutions of the Dirac equation in an axisymmetric black hole geometry
\emph{Commun. Pure Appl. Math.}  {\bf 53} 902,
arXiv: gr-qc/9905047


\bibitem{Finster:2000jz}Finster F, Kamran N, Smoller J and Yau S T 2003
The long time dynamics of Dirac particles in the Kerr--Newman black hole geometry
\emph{Adv. Theor. Math. Phys.} {\bf 7} 25,
arXiv: gr-qc/0005088


\bibitem{DG}Dolan S and Gair J 2009 
The massive Dirac field on a rotating black hole spacetime: angular solutions
\emph{Class. Quantum Grav.} \textbf{26} 175020, 
arXiv:0905.2974 [gr-qc]


\bibitem{Roken:2015fja}Roken C
The Massive Dirac Equation in Kerr Geometry: 
Separability in Eddington-Finkelstein-Type Coordinates and Asymptotics,
arXiv:1506.08038 [gr-qc]


\bibitem{Unruh}Unruh W G 1974
Second quantization in the Kerr metric
\emph{Phys. Rev.} D {\bf 10} 3194 



\bibitem{Teukolsky}Teukolsky S A 1973 
Perturbations of a rotating black hole I. Fundamental equations for gravitational, electromagnetic, and neutrino-field perturbations
\emph{Astrophys. J.} {\bf 185} 635


\bibitem{RWPU}Richartz M, Weinfurtner S, Penner A J and Unruh W G 2009
Generalised superradiant scattering
\emph{Phys. Rev.} D {\bf 80} 124016, 
arXiv:0909.2317 [gr-qc]



\bibitem{Mazur}Mazur P O 1982
Proof of uniqueness of the Kerr--Newman black hole solution 
\emph{J. Math. Phys.} \textbf{15} 3173 


\bibitem{Bunting}G. L. Bunting, 
Proof of the uniqueness conjecture for black holes, 
1983, Ph.D. Thesis, University of New England, Armidale, Australia 


\bibitem{WY}Wu T T and Yang C N 1976 
Dirac monopole without strings: monopole harmonics
\emph{Nucl. Phys.} B \textbf{107} 365


\bibitem{Semiz3}Semiz I 1992 
The Dirac equation is separable on the dyon black hole metric 
\emph{Phys. Rev.} D {\bf 46} 5414,
arXiv: gr-qc/9207010

\bibitem{Carter}Carter B 1968 
Hamilton--Jacobi and Schrodinger separable solutions of Einstein's equations 
\emph{Commun. Math. Phys.} {\bf 10} 280 





\bibitem{Hubeny}Hubeny V E 1999 
Overcharging a black hole and cosmic censorship 
\emph{Phys. Rev.} D \textbf{59} 064013, 
arXiv: gr-qc/9808043

\bibitem{FY}de Felice F and Yunqiang Yu 2001
Turning a black hole into a naked singularity 
\emph{Class. Quantum Grav.} \textbf{18} 1235 



\bibitem{JS}Jacobson T and Sotiriou T P 2009 
Over-spinning a black hole with a test body 
\emph{Phys. Rev. Lett.} \textbf{103} 141101; 
\emph{Phys. Rev. Lett.} \textbf{103} (2009) 209903 (Erratum),
arXiv:0907.4146 [gr-qc]

\bibitem{Jensen}Jensen B 1995 
Stability of black hole event horizons
\emph{Phys. Rev.} D \textbf{51} 5511,
arXiv: gr-qc/9408020

\bibitem{CG}Cohen J M and Gautreau R 1979 
Naked singularities, event horizons, and charged particles
\emph{Phys. Rev.} D \textbf{19} 2273


\bibitem{BR}Bekenstein J D and Rosenzweig C 1994
Stability of the black hole horizon and the Landau ghost
\emph{Phys. Rev.} D \textbf{50} 7239,
arXiv: gr-qc/9406024


\bibitem{MS}Matsas G E A and da Silva A R R 2007 
Overspinning a nearly extreme charged black hole via a quantum tunneling process 
\emph{Phys. Rev. Lett.} \textbf{99} 181301,
arXiv:0706.3198 [gr-qc]

\bibitem{MRSSV}Matsas G E A, Richartz M, Saa A, da Silva A R R and Vanzella D A T 2009
Can quantum mechanics fool the cosmic censor?
\emph{Phys. Rev.} D \textbf{79} 101502(R), 
arXiv:0905.1077 [gr-qc]

\bibitem{RS2}Richartz M and Saa A 2011 
Challenging the weak cosmic censorship conjecture with charged quantum particles 
\emph{Phys. Rev.} D \textbf{84} 104021,
arXiv:1109.3364 [gr-qc]


\bibitem{RS}Richartz M and Saa A 2008
Overspinning a nearly extreme black hole and the Weak Cosmic Censorship conjecture
\emph{Phys. Rev.} D {\bf 78} 081503,
arXiv:0804.3921 [gr-qc]


\bibitem{Hod2}Hod S 2008 
Return of the quantum cosmic censor 
\emph{Phys. Lett.} B \textbf{668} 346,
arXiv:0810.0079 [gr-qc]



\bibitem{Hod:2000bi}Hod S and Piran T 2000
Cosmic censorship: The Role of quantum gravity
\emph{Gen. Rel. Grav.}  {\bf 32} 2333,
arXiv: gr-qc/0011003

\bibitem{Hod:2002pm}Hod S 2002
Cosmic censorship, area theorem, and selfenergy of particles
\emph{Phys. Rev.} D {\bf 66} 024016,
arXiv: gr-qc/0205005

\bibitem{Hod:2008zza}Hod S 2008
Weak Cosmic Censorship: As strong as ever
\emph{Phys. Rev. Lett.}  {\bf 100} 121101,
arXiv:0805.3873 [gr-qc]

\bibitem{Hod:1999kn}Hod S 1999
Black hole polarization and cosmic censorship
\emph{Phys. Rev.} D {\bf 60} 104031,
arXiv: gr-qc/9907001

\bibitem{Hod:2013vj}Hod S 2013
Cosmic censorship: formation of a shielding horizon around a fragile horizon
\emph{Phys. Rev.} D {\bf 87} 2 024037,
arXiv:1302.6658 [gr-qc]


\bibitem{GL}Gao S and Lemos J P S 2008 
Collapsing and static thin massive charged dust shells in a 
Reissner--Nordstr\"om black hole background in higher dimensions
\emph{Int. J. Mod. Phys.} A \textbf{23} 2943,
arXiv:0804.0295 [hep-th]

\bibitem{Rocha:2011wp}Rocha J V and Cardoso V 2011
Gravitational perturbation of the BTZ black hole induced by test particles and weak cosmic censorship in AdS spacetime
\emph{Phys. Rev.} D {\bf 83} 104037,
arXiv:1102.4352 [gr-qc]

\bibitem{Rocha:2014jma}Rocha J V and Santarelli R 2014
Flowing along the edge: spinning up black holes in AdS spacetimes with test particles
\emph{Phys. Rev.} D {\bf 89} 6  064065,
arXiv:1402.4840 [gr-qc]

\bibitem{Delsate:2014iia}Delsate T, Rocha J V and Santarelli R 2014
Collapsing thin shells with rotation
\emph{Phys. Rev.} D {\bf 89} 121501,
arXiv:1405.1433 [gr-qc]

\bibitem{Rocha:2015tda}Rocha J V 2015
Gravitational collapse with rotating thin shells and cosmic censorship
\emph{Int. J. Mod. Phys.} D {\bf 24} 09  1542002,
arXiv:1501.06724 [gr-qc]


\bibitem{BCNR}Bouhmadi-Lopez M, Cardoso V, Nerozzi A and Rocha J V 2010 
Black holes die hard: can one spin up a black hole past extremality?
\emph{Phys. Rev.} D \textbf{81} 084051, 
arXiv:1003.4295 [gr-qc]

\bibitem{BCK1}Barausse E, Cardoso V and Khanna G 2010 
Test bodies and naked singularities: is the self-force the cosmic censor?
\emph{Phys. Rev. Lett.} \textbf{105} 261102,
arXiv:1008.5159 [gr-qc]

\bibitem{BCK2}Barausse E, Cardoso V and Khanna G 2011 
Testing the Cosmic Censorship Conjecture with point particles: 
the effect of radiation reaction and the self-force
\emph{Phys. Rev.} D \textbf{84} 104006,
arXiv:1106.1692 [gr-qc]

\bibitem{IST}Isoyama S, Sago N and Tanaka T 2011 
Cosmic censorship in overcharging a Reissner--Nordstr\"om black hole via charged particle absorption
\emph{Phys. Rev.} D \textbf{84} 124024,
arXiv:1108.6207 [gr-qc]


\bibitem{SS}Saa A and Santarelli R 2011 
Destroying a near-extremal Kerr--Newman black hole 
\emph{Phys. Rev.} D \textbf{84} 027501,
arXiv:1105.3950 [gr-qc]


\bibitem{ZG}Zhang Y and S. Gao S 2014
Testing cosmic censorship conjecture near extremal black holes with cosmological constants
\emph{Int. J. Mod. Phys.} D {\bf 23} 1450044,
arXiv:1309.2027 [gr-qc]

\bibitem{LB}Li Z and Bambi C 2013
Destroying the event horizon of regular black holes
\emph{Phys. Rev.} D {\bf 87} 12 124022,
arXiv:1304.6592 [gr-qc]

\bibitem{GZ}Gao S and Zhang Y 2013
Destroying extremal Kerr--Newman black holes with test particles
\emph{Phys. Rev.} D {\bf 87} 4 044028,
arXiv:1211.2631 [gr-qc]


\bibitem{CB}Colleoni M and Barack L 2015 
Overspinning a Kerr black hole: the effect of self-force
\emph{Phys. Rev.} D {\bf 91} 104024,
arXiv:1501.07330 [gr-qc]

\bibitem{Colleoni:2015ena}Colleoni M, Barack L, Shah A G and van de Meent M 2015
Self-force as a cosmic censor in the Kerr overspinning problem
\emph{Phys. Rev.} D {\bf 92} 8 084044,
arXiv:1508.04031 [gr-qc]


\bibitem{Semiz:2015pna}Semiz I and Duztas K 2015
Weak Cosmic Censorship, Superradiance and Quantum Particle Creation
\emph{Phys. Rev.} D {\bf 92} 10 104021,
arXiv:1507.03744 [gr-qc]


\bibitem{Duztas:2015oja}Duztas K and Semiz I
Black hole evaporation as a Cosmic Censor,
arXiv:1508.06685 [gr-qc]

\bibitem{Duztas:2015qqa}Duztas K
Absorption probability of neutrino fields and Hawking radiation,
arXiv:1503.05061 [gr-qc]



\bibitem{SemizKG}Semiz I 1992 
Klein-Gordon equation is separable on dyon black hole metric
\emph{Phys. Rev.} D {\bf 45} 532



\bibitem{PQR}Pugliese D, Quevedo H and Ruffini R 2013
Equatorial circular orbits of neutral test particles in the Kerr--Newman spacetime
\emph{Phys. Rev.} D {\bf 88} 2  024042,
arXiv:1303.6250 [gr-qc]


\bibitem{KMB}Kong L, Malafarina D and Bambi C 2014
Can we observationally test the weak cosmic censorship conjecture?
\emph{Eur. Phys. J.} C {\bf 74} 2983,
arXiv:1310.8376 [gr-qc]


\bibitem{VE}Virbhadra K S and Ellis G F R 2002 
Gravitational lensing by naked singularities 
\emph{Phys. Rev.} D \textbf{65} 103004


\bibitem{VK}Virbhadra K S and Keeton C R 2008 
Time delay and magnification centroid due to gravitational lensing
by black holes and naked singularities 
\emph{Phys. Rev.} D \textbf{77} 124014,
arXiv:0710.2333 [gr-qc]

\bibitem{Joshi:2013dva}Joshi P S, Malafarina D and Narayan R 2014
Distinguishing black holes from naked singularities through their accretion disc properties
\emph{Class. Quantum Grav.}  \textbf{31} 015002,
arXiv:1304.7331 [gr-qc]

\bibitem{Ortiz:2015rma}Ortiz N, Sarbach O and Zannias T 2015
The shadow of a naked singularity
\emph{Phys. Rev.} D {\bf 92} 4 044035,
arXiv:1505.07017 [gr-qc]



\bibitem{W}Weinberg S 1972 
\emph{Gravitation and Cosmology: Principles and Applications of the General Theory of Relativity}
(New York: Wiley) Chapter 12 Section 5 

\bibitem{PS}Peskin M E and Schroeder D V 1995 
\emph{An Introduction to Quantum Field Theory}
(Reading: Addison-Wesley)


\end{thebibliography}
\end{document}